\documentclass[prl,twocolumn,showpacs,superscriptaddress,floatfix,footinbib,amsmath,amssymb]{revtex4}

\usepackage[utf8]{inputenc}
\usepackage{graphicx}
\usepackage[normalem]{ulem}
\usepackage[usenames,dvipsnames]{color}
\usepackage[nointegrals]{wasysym}
\usepackage{commands}
\usepackage{cleveref}
\usepackage{upgreek}
\usepackage{amsmath}

\begin{document}


\title{Electrostatic potential shape of gate defined quantum point contacts}

\author{M. Geier}\affiliationFUtheory
\author{J. Freudenfeld}\affiliationPDI
\author{J. T. Silva}\affiliationPDI
\author{V. Umansky}\affiliationWIS
\author{D. Reuter\footnote{Present address: Department of Physics, Paderborn University, Warburger Straße 100, 33098 Paderborn}}\affiliationBochum
\author{A. D. Wieck}\affiliationBochum
\author{P. W. Brouwer}\affiliationFUtheory
\author{S. Ludwig}\affiliationPDI

\date{\today}

\pacs{
}
\begin{abstract}
Quantum point contacts (QPC) are fundamental building blocks of nanoelectronic circuits. For their emission dynamics as well as for interaction effects such as the 0.7-anomaly the details of the electrostatic potential are important, but the precise potential shapes are usually unknown. Here, we measure the one-dimensional subband spacings of various QPCs as a function of their conductance and compare our findings with models of lateral parabolic versus hard wall confinement. We find that a gate-defined QPC near pinch-off is compatible with the parabolic saddle point scenario. However, as the number of populated subbands is increased Coulomb screening flattens the potential bottom and a description in terms of a finite hard wall potential becomes more realistic.
\end{abstract}

\maketitle

\section{introduction}

Given the importance of quantum point contacts (QPC) as fundamental building blocks of nanoelectronic circuits and the vast amount of literature about them \cite{Landauer1981,QPCsBeenakker1988,QPCsWharam1988,Berggren2010,Micolich2011}, surprisingly little is known about the shape of their electrostatic potential as a function of gate voltages. However, knowledge of the precise confinement potential is crucial for understanding interaction effects in QPCs \cite{Rejec2006,Koop2007,Bauer2013,Heyder2015} as well as their carrier emission dynamics \cite{Topinka2000,Freudenfeld2020-1}, which is central for optimizing a quantum electronic circuit. The lateral confinement defines the mode structure of the one-dimensional (1D) channel while the longitudinal potential shape governs the coupling of the 1D modes into the surrounding 2DES. Populating the 1D channel with electrons by increasing the voltage applied to the split gates enhances Coulomb screening inside the constriction. As a consequence, the lateral confinement potential undergoes a transition from an unscreened approximately parabolic shape near pinch-off towards a screened potential for many occupied 1D subbands. Such a transition had been theoretically predicted \cite{Laux1988}. Here, we experimentally demonstrate it using transport spectroscopy at finite source-drain voltage.

Details of the confinement vary between individual devices produced by various layouts based on different methods, which include the field effect \cite{QPCsWharam1988,QPCsBeenakker1988}, etching \cite{Martin2008} or oxidation \cite{Senz2001} techniques and more \cite{Fricke2006,Roessler2010}. The manifestation of 1D conductance quantization, $G=N\gq$ with $\gq=2e^2/h$ and $N=1,2,3,\dots$, at cryogenic temperatures is often seen as a quality feature of QPCs. An ``optimally'' designed QPC has several conductance steps that are approximately equidistant in gate voltage as the QPC is opened up starting from pinch-off at $G=0$. It is tempting to interpret the presence of equidistant conductance steps \cite{QPCsBeenakker1988,Taboryski1995,Thomas1998,Hew2008,Roessler2011,Burke2012} as a signature of a parabolic transverse confinement potential as introduced in Ref.\ \cite{Buettiker1990}, since such a potential has transverse modes at equally spaced energies. However, this interpretation is questionable as the distance of the conductance steps as a function of gate voltage is not one-to-one related with the energy spacing of the 1D modes \cite{Roessler2011,Micolich2011-1}. 

\begin{figure*}[tb]
\includegraphics[width=1.9\columnwidth]{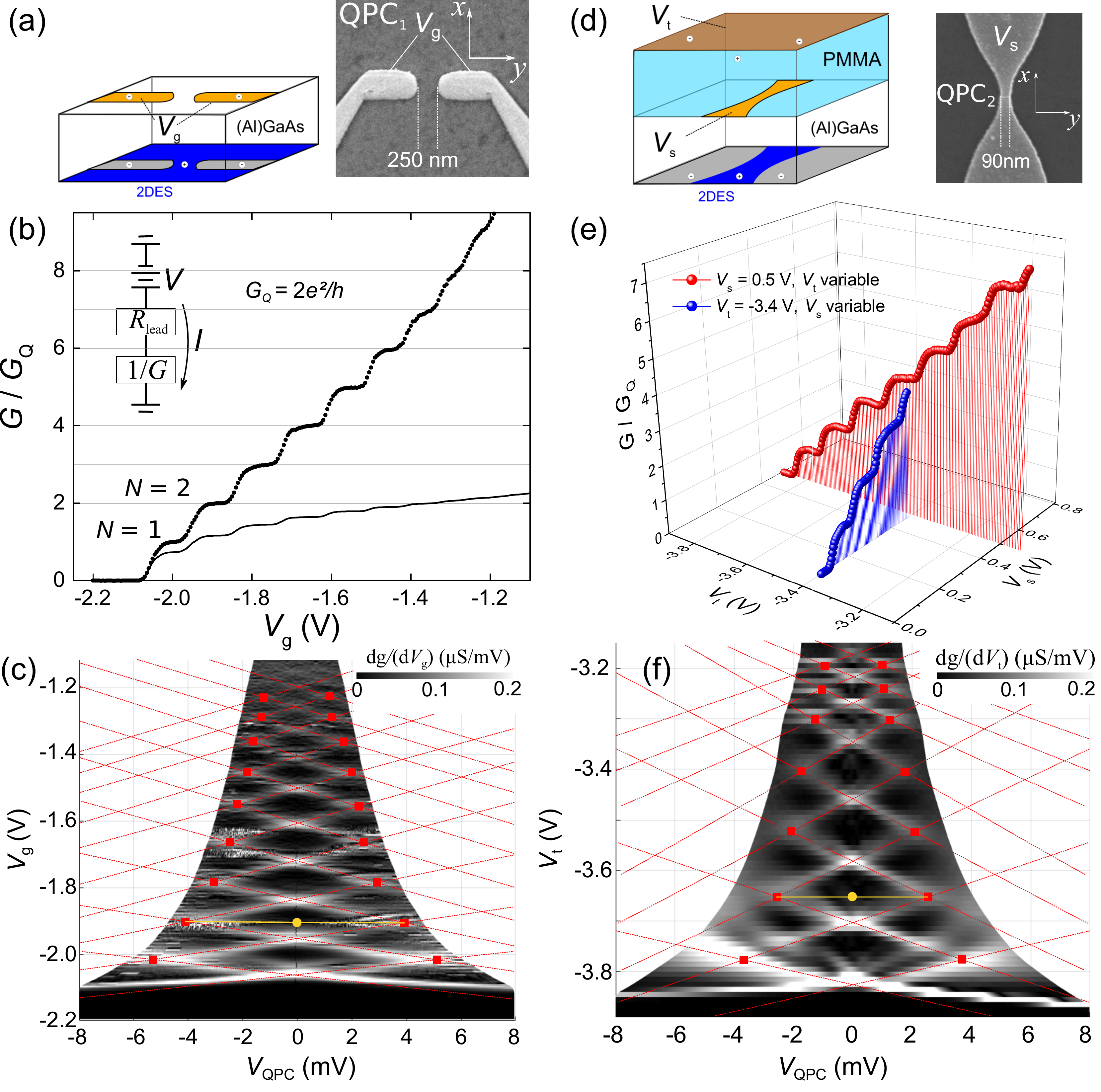}
\caption{(a) Scanning electron microscope (SEM) picture of Ti/Au gates (light gray) on the wafer surface (dark) of QPC$_1$ and sketch of the electric field effect device. Negative voltage \vg\ applied to the gates (yellow) is used to locally deplete the 2DES (blue, where conducting, gray where depleted) 107\,nm beneath the surface.
(b) Pinch-off curve $G(\vg)/\gq$ of QPC$_1$ using the source-drain voltage $V=-0.5$\,mV; solid line: raw data; dots: corrected for lead resistance $R_\text{lead}=4.62$\,k$\Omega$ which includes $4.4$\,k$\Omega$ resistance of external RC filters; inset: simplified circuit diagram of the measurement.
(c) Finite bias spectroscopy of QPC$_1$, $\text d g/\text d \vg (V_\text{QPC},\vg)$, accounting for the lead resistance (see main text). Local maxima of $\text d g/\text d \vg$ (white lines) indicate transitions between adjacent conductance plateaus.
(d) SEM picture of a screen gate equivalent to that of QPC$_2$. As shown in the sketch, the actual device is covered with a 130\,nm thick layer of cross-linked PMMA which carries a global top gate.
(e) Pinch-off curves of QPC$_2$ corrected for a gate voltage dependent lead resistance, including a constant $4.4$\,k$\Omega$ resistance of external RC filters, cf. Fig. \ref{fig:lead_resistances}{}: $G(\vt)/\gq$ for $\vs=0.5\,$V and $G(\vs)/\gq$ for $\vt=-3.4\,$V at $V=-0.1$\,mV.
(f) $\text d g/\text d \vg (V_\text{QPC},\vt)$ of QPC$_2$, accounting for the lead resistance.
Additional lines and symbols in panels (c) and (f) are explained in the main text.}
\label{fig:qpcs}
\end{figure*}
We study QPCs of two different designs, but both defined using gate voltages by means of the electric field effect. In agreement with previous publications \cite{Buettiker1990,Berggren2010,Burke2012,Heyder2015} our findings are consistent with a parabolic confinement potential near the pinch-off point of the QPCs. However, as the conductance of a QPC is increased, more and more carriers populate the 1D subbands and thereby arrange themselves to partially screen the electric field induced by the applied gate voltages. The resulting effective potential is then a function of the position of all charges, which also includes the usually not well-known distribution of surface states and charged bulk defects. A precise theoretical description of this screening effect requires a three-dimensional self-consistent calculation solving the classical Poisson equations together with the quantum mechanical Schr\"odinger equations \cite{Laux1988,Yakimenko2013,nextnano}. A self consistent Poisson-Schr\"odinger calculation performed for a set of fictitious boundary conditions and for the case of a standard split-gate defined QPC suggests a transition from a parabolic lateral confinement for $N=1$ towards a truncated parabola and, eventually, a hard wall confinement as $N$ is increased \cite{Laux1988}. 
To test this scenario we measure non-linear response transport through our QPC from which we identify the energy spacings between its highest occupied 1D modes. We compare our results to the two extreme scenarios for the lateral electrostatic confinement: parabolic confinement as, e.g., in Refs.\ \cite{Weisz1989,Hew2008,Song2009,Roessler2011} and a hard-wall confinement as, e.g., in Refs.\ \cite{vanWees1988,Gloos2006}. Our results are inconsistent with parabolic confinement for $N\ge4$ but are consistent with a transition from a parabolic lateral confinement at $N=1$ towards a hard wall potential as the QPCs are opened up.

\section{Transport spectroscopy of quantum point contacts}

Our QPCs are formed using the electric field effect in a 2D electron system (2DES) embedded 107\,nm beneath the surface of a (Al,Ga)As/GaAs heterostructure. The 2DESs Fermi energy and mobility measured at cryogenic temperatures are $\ef\simeq10.9$\,meV and $\mu_e\simeq2.6\times10^6\,\text{cm}^2/$Vs for QPC$_1$ and similar for QPC$_2$. We performed all measurements in a helium-3 evaporation cryostat at temperatures near $T=250$\,mK. In \fig{fig:qpcs}a and (d) 
we present scanning electron microscope images of the two QPC samples and sketches of the gate layouts. For QPC$_1$ shown in panel (a) we use a standard split-gate layout and define the 1D constriction of the 2DES by applying a negative voltage $V_\text g$ to both gates while the 2DES and a back gate approximately 500\,$\mu$m below the surface are at ground potential. The resulting linear response pinch-off curve, $G(\vg)/\gq$, is presented in \fig{fig:qpcs}b. It features clear and, for $N<6$, nearly equidistant steps of quantized conductance. To create the second QPC$_2$, see Fig.\ \ref{fig:qpcs} (d), we use a global top gate to globally deplete the 2DES. Only below a screen gate placed in between the top gate and the 2DES we induce a finite density of free electrons \cite{Bachsoliani2017}. The screen gate shapes a narrow constriction, i.e., a QPC between 2D leads. Both, the QPC conductance and the carrier density in the leads are controlled by the combination of the voltages \vt\ and \vs\ applied to the top gate and screen gate, respectively. We present example pinch-off curves $G(\vt)$ for fixed $\vs=0.5\,$V and $G(\vs)$ for constant $\vt=-3.4\,$V in \fig{fig:qpcs}e. Note, that the screen gate voltage is restricted to $\vs\lesssim0.5\,$V as larger \vs\ causes a leakage current from the gate into the 2DES (as expected for a Schottky barrier).

All our pinch-off curves feature smooth transitions between quantized conductance plateaus. They indicate that the potential varies slowly and smoothly in current direction, reminiscent of a parabolic potential profile in the longitudinal direction, which results in reflectionless contacts between constriction and leads.

Quantized conductance is a consequence of the energy quantization in a 1D channel caused by the lateral confinement of the constriction. To experimentally determine the energies of the 1D modes we need a known energy scale to compare with. For this reason we measure the differential conductance $g=\text dI/\text dV$ (e.g., using a lock-in amplifier) as a function of source-drain voltage $V$ along the pinch-off curves. In panels (c) and (f) of \fig{fig:qpcs}{} we plot the differential transconductances $\text dg/\text d\vg$  ($\text dg/\text d\vt$) for the two QPCs as a function of the gate voltage and the bias voltage \vqpc\ (defined below) dropping across the QPC. In these plots steps of the conductance $G(\vg,\vqpc)$ [$G(\vt,\vqpc)$] appear as lines of positive differential transconductance (white). Red lines are a guide for the eye, indicating resonances between the 1D modes and the chemical potentials of the source and drain leads. Along the $N$th line of positive (negative) slope counted from the bottom of the plot, the $N$th 1D subband bottom energy is equal to the chemical potential in the source (drain) lead, $\varepsilon_N=\mu_\text{S}$ ($\varepsilon_N=\mu_\text{D}$). The lines frame diamond shaped regions around $\vqpc=0$. Within these regions the conductance takes the quantized values $G=N\gq$. Intersection points at $\vqpc=0$ indicate steps of the linear response pinch-off curves, i.e., $G=(N-0.5)\gq$. At intersection points at finite $\vqpc\ne0$ the chemical potential drop across a QPC equals the energy spacing between the corresponding 1D modes,  $|\mu_\text S-\mu_\text D|=e\vqpc=\varepsilon_N-\varepsilon_M$.
The additional curved lines of enhanced differential transconductance within the $N=1$ diamond indicate the 0.7-anomaly \cite{Rejec2006,Koop2007,Micolich2011,Bauer2013,Heyder2015} which is not a topic of this article.

\begin{figure}[h]
\includegraphics[width=\columnwidth]{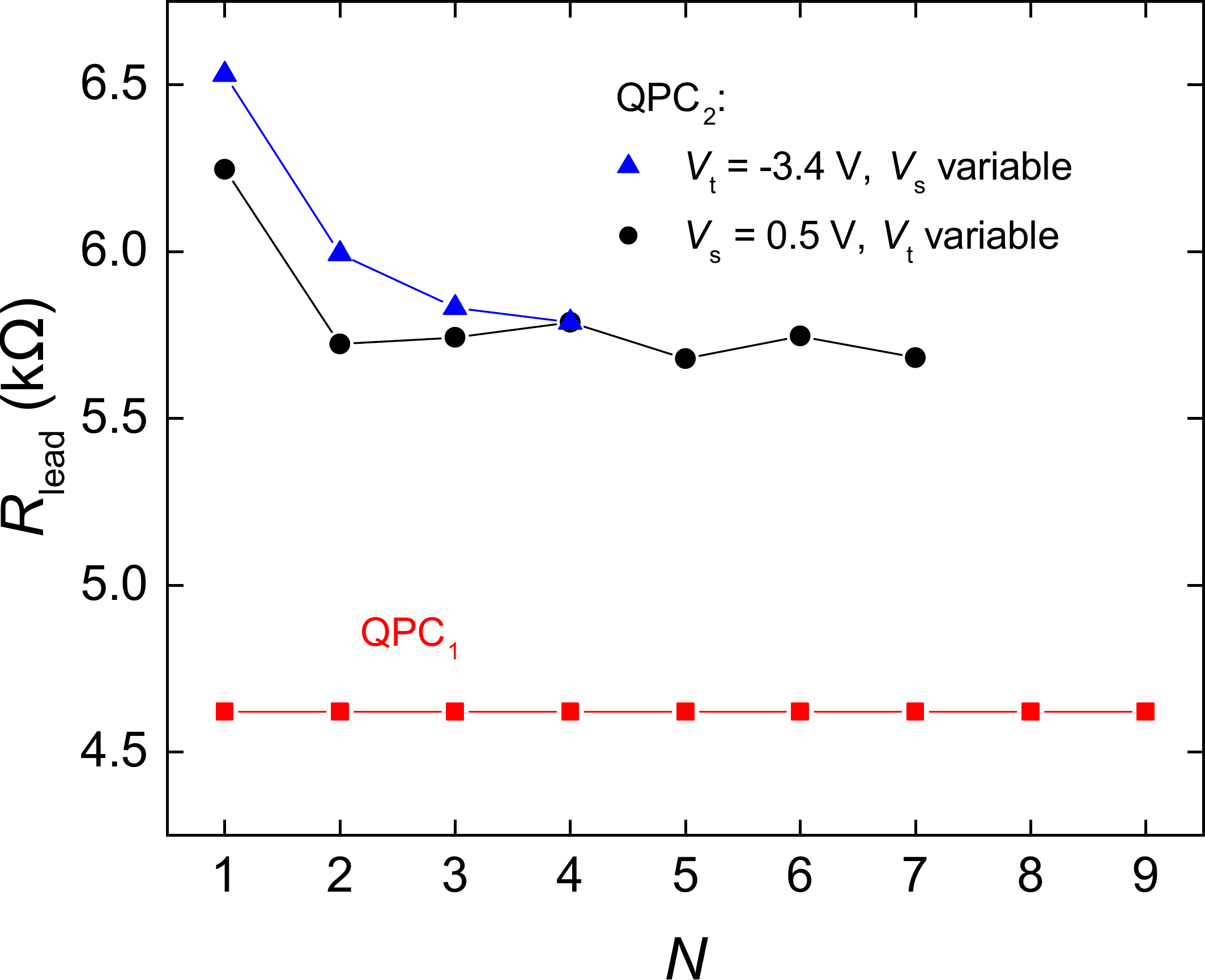}
\caption{Resistances $R_\text{lead}$ of the leads to the QPCs, cf.\ sketch in Fig.\ \ref{fig:qpcs}(b). For the split gate design of QPC$_1$ $R_\text{lead}$ is constant while it is a function of gate voltages for QPC$_2$.}
\label{fig:lead_resistances}
\end{figure}

Since the source-drain voltage $V$ is applied across the QPC \textit{and its leads} (which is always the case, because of the finite contact sizes even for a four-terminal measurement), the voltage drop across a QPC is $V_\text{QPC} = V - V_\text {lead} = V - R_\text{lead}I$, cf.\ sketch in \fig{fig:qpcs}b. The lead resistance can be directly determined from the linear response pinch-off curves by forcing the conductance plateaus to their quantized values, $R_\text{lead}=V/I-(N\gq)^{-1}$. Our pinch-off curves in panels (b) and (e) of \fig{fig:qpcs}{} are already corrected for the lead resistances, while for QPC$_1$ we additionally plot the uncorrected curve, i.e., the raw data as a solid line. For completeness we present the lead resistances for all three pinch-off curves in \fig{fig:lead_resistances}{}.
From these we determine the voltage drop across the QPC, $\vqpc=V/(R_\text{lead}G+1)$, which is the x-axis in panels (c) and (f) of \fig{fig:qpcs}{}. The tapered shape of the region of plotted data is a result of correcting for the lead resistances (we measured between $-8\text{mV}\le V\le8\,$mV).

\begin{figure}[h]
\includegraphics[width=0.9\columnwidth]{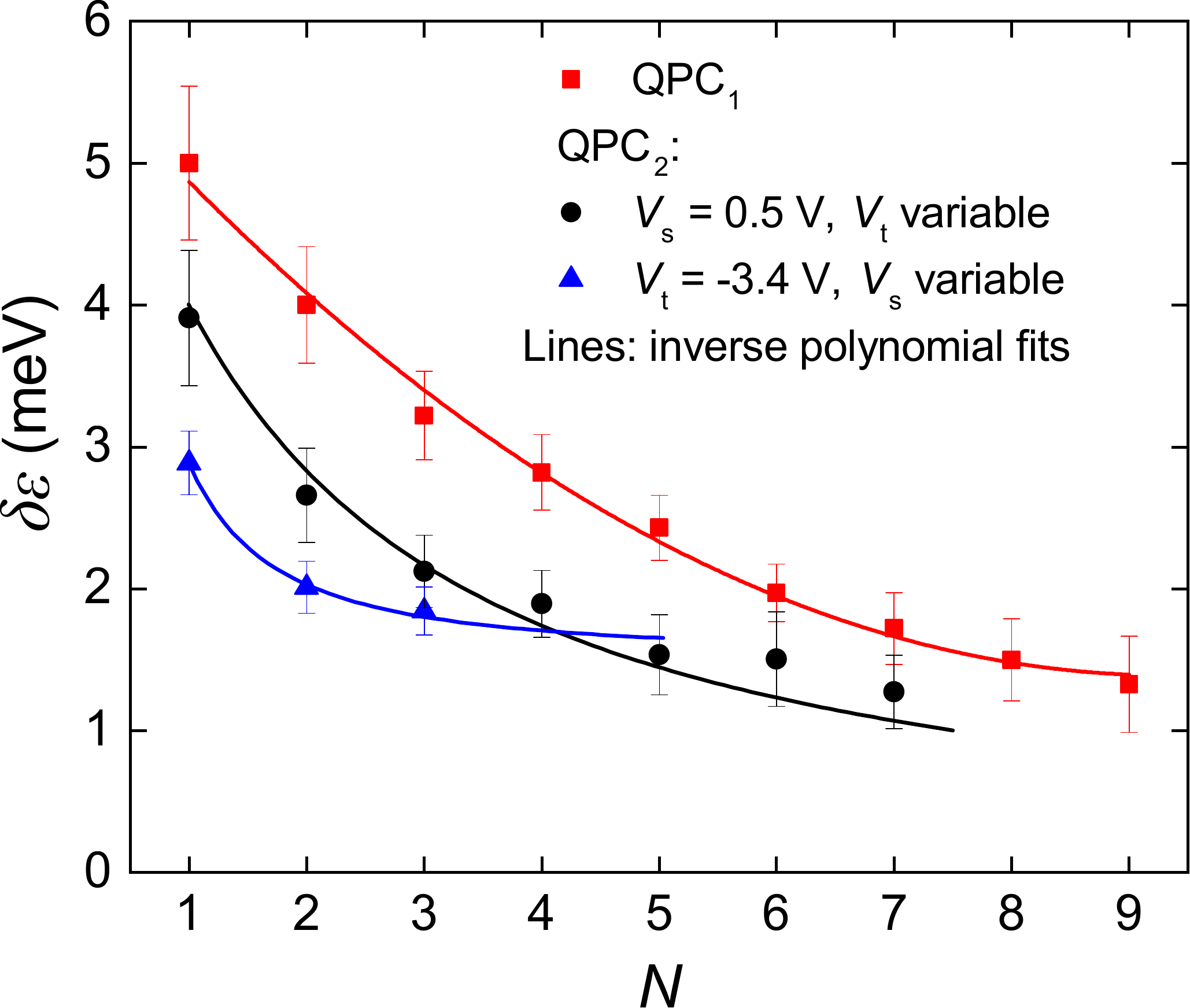}
\caption{Subband spacings $\delta\epsilon(N)$ of both QPCs for the three pinch-off curves presented in Fig.\ \ref{fig:qpcs}{}. Lines are guides for the eyes. At the intersection of the two lines of QPC$_2$ the gate voltages \vs\ and \vt\ would be identical for both measurements. Error bars reflect the uncertainties of the red lines in Figs.\ \ref{fig:qpcs}(c) and (d).
}
\label{fig:subbandspacings}
\end{figure}

At the intersection points marked by red squares in panels (c) and (f) of \fig{fig:qpcs}{} the bias \vqpc\ is precisely equal to the energy spacings between subsequent subbands,
\begin{align}
\label{eq:1}
  \delta\varepsilon(N)=&\,
  \varepsilon_{N+1}-\varepsilon_N
  \nonumber \\ =&\, e\vqpc.
\end{align} 
We plot $\delta\varepsilon(N)$ in \fig{fig:subbandspacings}{}
for all three pinch-off curves. Related to the variations in geometry the three implementations of QPCs have different subband spacings. However, as a general feature we observe a strong decrease of $\delta\varepsilon(N)$ as the QPCs are opened and $N$ is increased.

\section{Hard wall versus parabolic lateral confinement}

Given reflectionless contacts the conductance of a QPC is limited by its strongest lateral confinement in the center of the constriction. The measured subband spacings are uniquely related with this lateral confinement. In the following, we compare the two most common models describing the lateral confinement, namely a hard-wall versus a parabolic potential. These two models may be considered the extreme limits of a ``continuum'' of realistic scenarios for the transverse confinement.

\subsection{Lateral hard-wall potential} 

\begin{figure*}[bt]
\includegraphics[width=2\columnwidth]{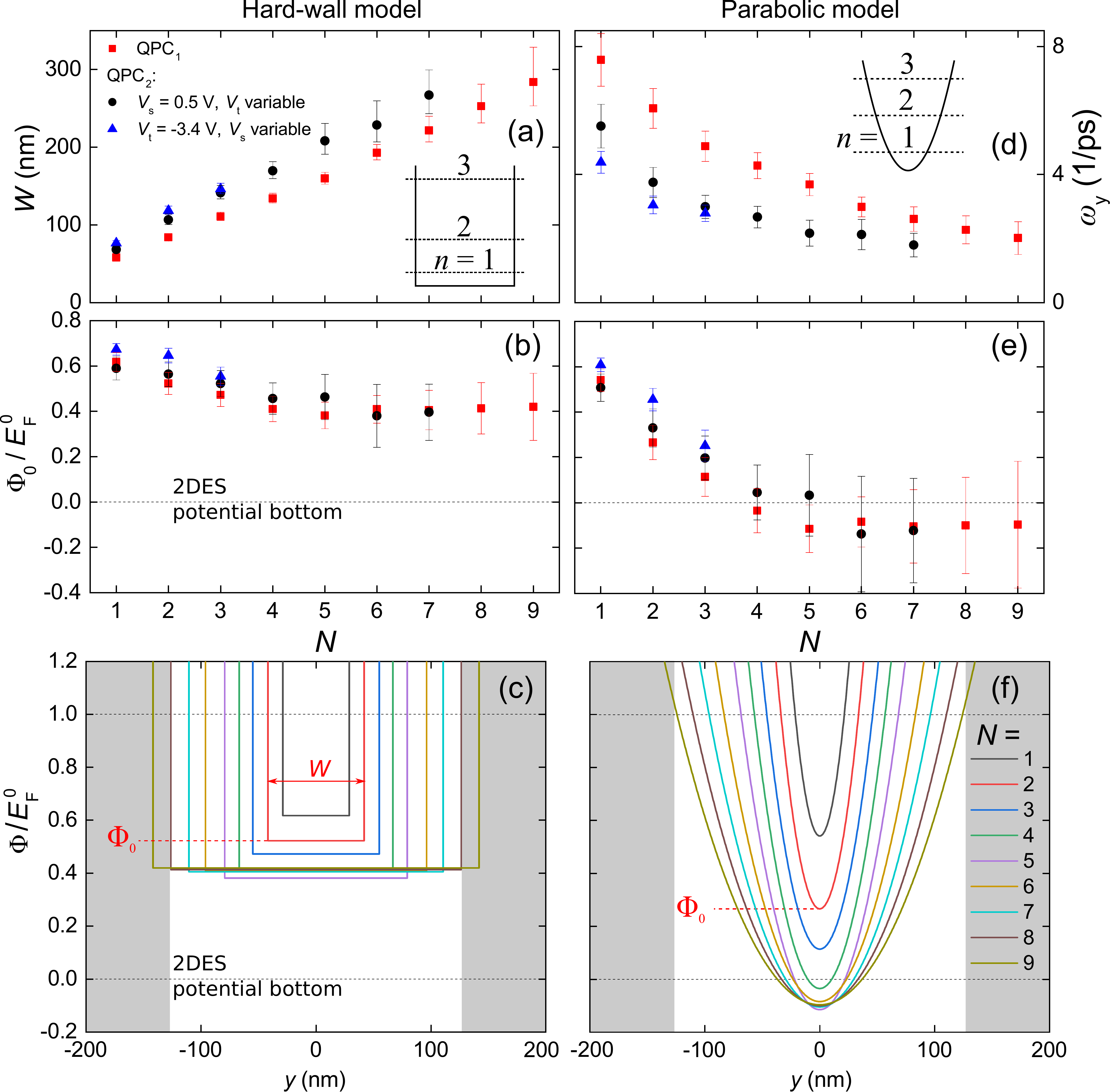}
\caption{Comparison between hard-wall (left column) and parabolic (right column) potential models of the lateral confinement. 
(a) Width of hard-wall potential $W(N)$.
(b) Offset of hard-wall potential $\Phi_0(N)$.
(c) Shape of hard-wall potential for $1\le N\le9$, only for QPC$_1$.
(d) Curvature of parabolic potential $\omega_y(N)$.
(e) Offset of parabolic potential $\Phi_0(N)$.
(f) Shape of parabolic potential for $1\le N\le9$, only for QPC$_1$.
Error bars in panels (a), (b), (d) and (e) are calculated by error propagation from the error of $\delta\varepsilon (N)$, cf.\ Fig.\ \ref{fig:subbandspacings}.
}
\label{fig:quantumwell-parameter}
\end{figure*}
For the lateral hard-wall potential we model the transverse confinement potential $\Phi(y)$ as
\begin{equation}
\Phi(y)= \begin{cases}
 \Phi_0, & |y| \leq {W}/{2} \\
 \infty, & |y| > {W}/{2}\,,
\end{cases} 
\label{eq:classical_slit}
\end{equation}
where the two parameters $W$ and $\Phi_0$ are the width and offset of the hard-wall potential well. An offset can be caused by a partial depletion of the constriction related to the incomplete screening in a semi conductor with a small carrier density. The threshold energies for the transverse modes are 
\begin{equation}
E_n = \frac{\pi^2 \hbar^2 n^2}{2 m^\star W^2} + \Phi_0\, \label{eq:E-hardwall}
\end{equation}
where $m^\star=0.067 m_0$ is the effective mass of the electrons in GaAs, $m_0$ being the free electron mass. Using Eq.\ (\ref{eq:1}) to relate the bias voltage at the intersection points marked by the red squares in \fig{fig:qpcs}c and (f) to the subband spacing $\delta \varepsilon(N) = E_{N+1} - E_{N}$,
we calculate the widths
\begin{equation}
  W(N) = \pi \hbar\sqrt{\frac{2 N + 1}{2m\delta\varepsilon(N)}}.
  \label{eq:width-hardwall}
\end{equation}
Neglecting additional screening effects from the applied bias voltage, these values of $W(N)$ apply everywhere along the (almost horizontal) lines connecting pairs of red squares, see the yellow lines for $N=2$ in \fig{fig:qpcs}c and (f). In particular, this allows us to extend our estimate of the width $W(N)$ to $\vqpc=0$, indicated for $N=2$ by the yellow dot in \fig{fig:qpcs}c and (f). Substituting $W$ in \eq{eq:E-hardwall} with $W(N)$ we then find the potential offset $\Phi_0$ using the relation $\ef\simeq E_N+0.5\delta\epsilon(N)$, which gives
\begin{equation}
\Phi_0(N)\simeq\ef-\delta\varepsilon(N)\,\left(\frac{N^2}{2N+1}+\frac 12\right)\,.
\end{equation}
The potential shift by $0.5\delta\epsilon(N)$ accounts for the difference between the $N$th subband bottom $E_N$ and the Fermi level \ef\ in the center of each diamond at $\vqpc=0$, assuming symmetric coupling between the 1D constriction and both leads. (The assumption of symmetric coupling is confirmed by the fact that the lines connecting pairs of red squares in \fig{fig:qpcs}c and (f) are almost horizontal.)

\subsection{Lateral parabolic potential}

To model a lateral parabolic potential we use 
\begin{equation}
\Phi(y) = \Phi_0 + \frac{m \omega_y^2 y^2}{2},
\label{eq:QPCpotential_sim}
\end{equation}
where $\omega_y$ and $\Phi_0$ are the characteristic frequency and offset of the parabolic potential well. In analogy to the analysis assuming hard-wall potentials we determine the two parameters from the measured subband spacings. At the intersection points indicated with red squares in \fig{fig:qpcs}c and (f) we find
\begin{equation}
\hbar \omega_y(N) = e\vqpc=\delta\varepsilon(N) \label{eq:omega_y-parabolic}
\end{equation}
and in the centers of the diamonds at $\vqpc=0$ in addition
\begin{equation}
\Phi_0(N) \simeq E_F - N \hbar \omega_y\,. \label{eq:offset_parabolic}
\end{equation}

\subsection{Comparison of the two potential shapes}

In \fig{fig:quantumwell-parameter}{}
we directly compare our results for the hard-wall potential shown in the left column and for assuming parabolic confinement plotted on the right hand side. We present the parameters $W$ and $\Phi_0$ as a function of the subband number $N$ for all three QPC implementations for the hard-wall potential in panels (a) and (b) and $\omega_y$ and $\Phi_0$ for the parabolic potential in panels (d) and (e).
The results are qualitatively similar for the various implementations of QPCs; the variations in $W$ or $\omega_y$ between QPCs indicate that the lateral confinement potential of QPC$_2$ is slightly wider compared to QPC$_1$. In panels (c) and (f) showing the actual potentials for QPC$_1$, for comparison we indicate the lithographic distance of 250\,nm between the gates seen in the inset of \fig{fig:qpcs}a. It corresponds to the white area between regions of gray background. The width of the hard-wall potential slightly exceeds the lithographic width for $N=9$. QPC$_1$ does not show further plateaus for $N>9$.

Comparing the two models a substantial difference is visible in $\Phi_0(N)$. While for $N=1$ the potential offset is similar for both models with $\Phi_0/\ef\simeq0.6$, in case of the hard-wall potential it slowly decreases to $\Phi_0/\ef\simeq0.4$ at $N=4$ and stays approximately constant at that level as the QPC is opened further. In contrast, the decrease of the offset $\Phi_0(N)$ of the parabolic potential with $N$ is much steeper, such that for $N\gtrsim4$ it moves below the bottom of the conduction band in the 2D leads, indicated as a dashed line at $\Phi=0$. We are not aware of a realistic mechanism that could lead to such an over-screening of the negative voltages applied to the control gates (\vg\ for QPC$_1$ or \vt\ for QPC$_2$).

\section{Discussion and summary}

The main result of our simple analysis starting from the measured subband spacings $\delta\varepsilon(N)$ is, that for $N\ge4$ we can exclude a parabolic lateral confinement potential for our QPCs. Based on a self-consistent calculation it has been suggested that the increasing population of the 1D constriction with $N$ as a QPC is opened up leads to an increased screening of the electric field originating from the charged control gates.  For a gate defined QPC this process can cause a transition from a parabolic confinement for the case of little screening, i.e., $N=1$ towards a truncated potential with a flat bottom at larger $N$ where many carriers populate the constriction \cite{Laux1988}. Our findings are in favor of such a scenario. The hard-wall potential presents a somewhat unrealistic extreme case of strong screening. Nevertheless, for $N\ge4$ it seems more realistic than the other extreme, namely the parabolic potential. The true shape of the lateral confinement potential of a QPC for $N\ge4$ likely lies between these two extremes, maybe close to a truncated parabola \cite{Laux1988,Wharam1989}, i.e., a parabola with a flat bottom identical to that of a hard-wall potential but with smoothly increasing side walls of constant curvature as the case for a parabola.


In summary, a parabolic saddle point potential is likely a realistic description of a QPC near pinch-off, although our measurement can also be explained with a hard-wall confinement in this regime. However, as the QPC is opened up beyond $N\simeq4$, the parabolic lateral confinement turns out to be a bad approximation. In this regime of enhanced screening a hard-wall potential is the better approximation.

\section{appendix}

\subsection{Coupling between control gates and the QPC}

\begin{figure}
\includegraphics[width=1.0\columnwidth]{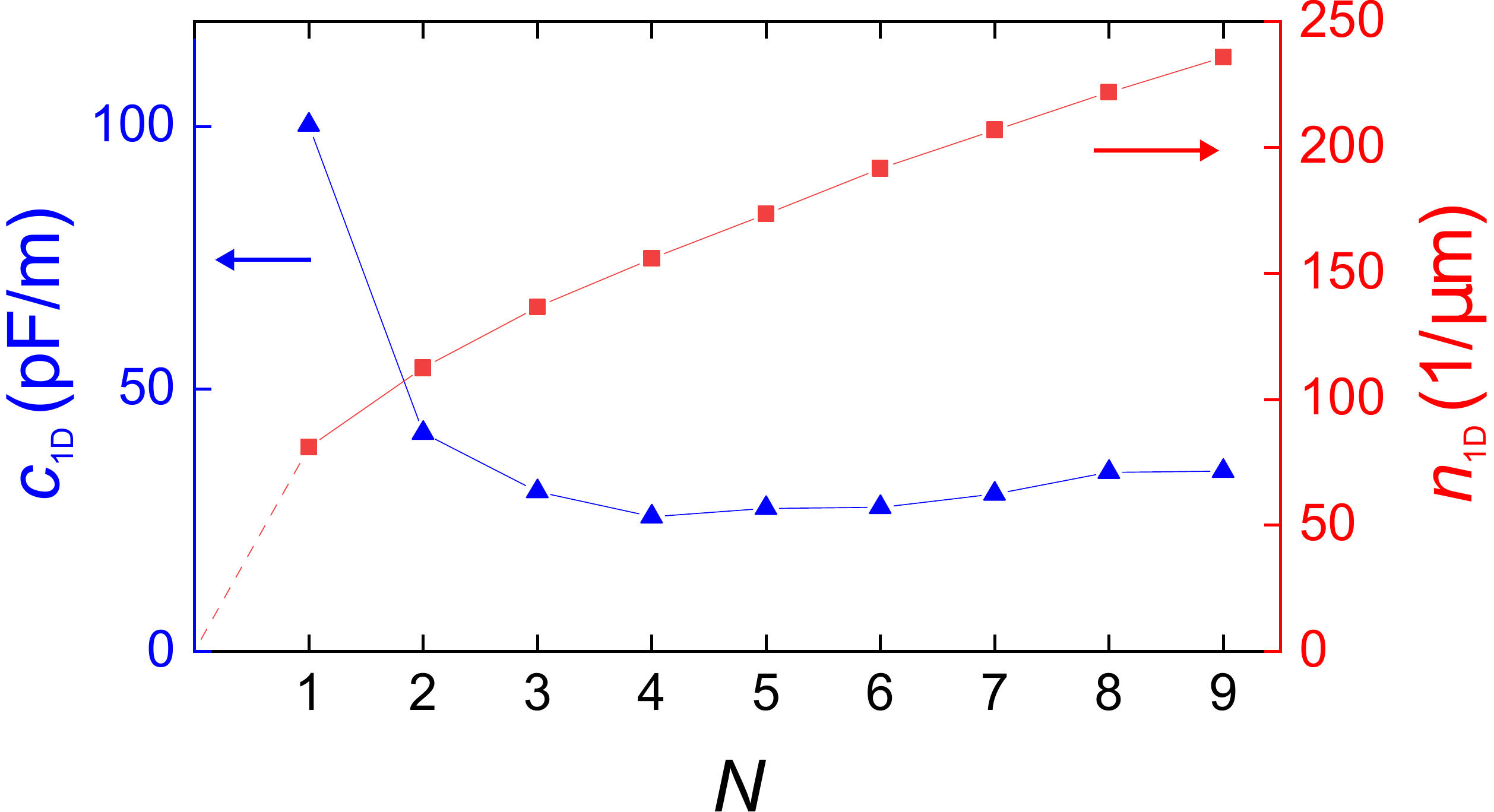}
\caption{1D carrier density $n_\text{1D}(N)$ assuming an infinitely long hard-wall 1D channel of width $W(N)$ and depth $\ef-\Phi_0(N)$ of QPC$_1$ (red squares, rhs axis) and corresponding 1D capacitance density $c_\text{1D}(N)$ (blue triangles, lhs axis).
}
\label{fig:capacitances}
\end{figure}

The electrostatic potential shaping the QPCs is generated and controlled via the field effect by applying voltages to nearby metal gates. The size of the plateaus of quantized conductance in the pinch-off curves as a function of gate voltage, cf.\ \fig{fig:qpcs}b and (e), is proportional to the capacitive coupling between the control gates and the QPC, which we approximate as a conducting 1D-channel with the carrier density $n_\text{1D}$. We determine the approximate capacitance per unit length between gate and QPC as
\begin{equation} \label{eq:capacity}
c_\text{1D}=e\delta n_\text{1D} / \delta V_\text{gate}\,,
\end{equation}
where $\delta n_\text{1D}$ is the carrier density increase as the voltage on the control gate is increased by $\delta V_\text{gate}$. If we take for  $\delta V_\text{gate}$ the voltage difference between two subsequent intersection points of the source- and drain-resonances at $\vqpc=0$ in \fig{fig:qpcs}c and (f), $\delta n_\text{1D}$ corresponds to the difference of the values of $n_\text{1D}$ at these points with $N$ versus $N+1$ subbands being populated. The 1D carrier density is 
\begin{equation} \label{eq:1Dcarrierdensity}
n_\text{1D}(N)
=\int_0^\infty D_\text{1D}(E)f(E)\text dE\,,
\end{equation}
where  $D_\text{1D}=\frac{1}{\pi\hbar}\sqrt{\frac{2m^\star}{E}}$ is the 1D electron density of states and $f(E)$ the Fermi-Dirac distribution. Given $\kb T\ll\ef$ we approximate $f(E)=1$ for $E<\ef$ and $f(E)=0$ for $E>\ef$. Summing up all 1D modes which are actually populated for the QPC tuned to the conductance $G=N\gq$ we find
\begin{align} \label{eq:1Dcarrierdens}
n_\text{1D}(N)
&=\frac{\sqrt{2m^\star }}{\pi \hbar} \sum_{n=1}^{N} \int_{E_n}^{E_\text F}\frac{1}{\sqrt{E-E_n}}dE\nonumber\\
&=\frac{\sqrt{8m^\star }}{\pi \hbar} \sum_{n=1}^{N} \sqrt{\ef-E_n}\,.
\end{align}
Inserting $\delta n_\text{1D}(N) = n_\text{1D}(N+1)-n_\text{1D}(N)$ from \eq{eq:1Dcarrierdens} in \eq{eq:capacity} we finally determine the 1D capacitance density as
\begin{equation} \label{eq:1Dcapacitance}
c_\text{1D}(N)=\frac{\sqrt{8m^\star e^2}}{\pi \hbar}\,\frac{\sqrt{\ef-E_{N+1}}}{\delta V_\text{gate}(N)}\,,
\end{equation}
where $\delta V_\text{gate}(N)$ is the width of the $Nth$ plateau of the pinch-off curve, cf.\ \fig{fig:qpcs}{}, measured between the conductance $(N+0.5)\gq$ and  $(N-0.5)\gq$. Substituting $E_{N+1}$ with the according eigen-energy of the hard-wall potential using \eq{eq:E-hardwall} we can now determine $c_\text{1D}(N)$. In \fig{fig:capacitances}{}
we present the 1D capacitance density $c_\text{1D}(N)$, which is the slope of the also shown 1D carrier density $n_\text{1D}(N)$. The strong decrease of the capacitance with $N$ for $N\le4$ is a direct signature of the increase of the screening of the electric field of the gates with growing carrier density.

In addition, the variations in capacitance as a function of $N$ explain the counter-intuitive result that the subband spacings $\delta\epsilon(N)$ strongly vary in a region of almost equal widths of the plateaus of quantized conductance of the pinch-off curve, cf.\ \twofigs{fig:qpcs}be and \fig{fig:subbandspacings}{}.

\subsection{Width of the 1D constriction as a function of gate voltage}

\begin{figure}
\includegraphics[width=1.0\columnwidth]{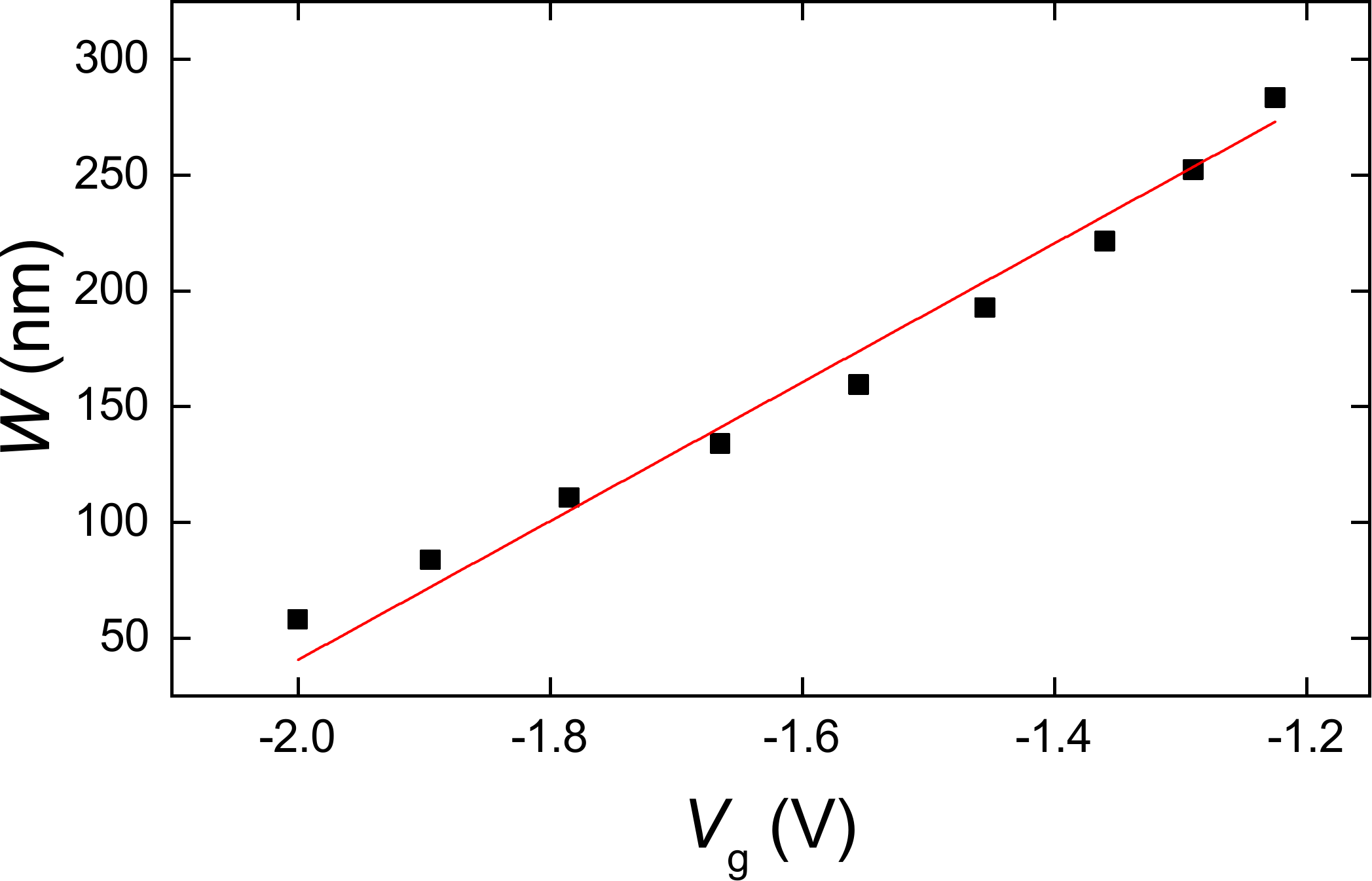}
\caption{Width of the hard-wall potential $W(\vg)$ for QPC$_1$ [same data as the $W(N)$ in Fig.\ \ref{fig:quantumwell-parameter}(a)]. The slope of the red line is $\text d W/\text d V_\text g=300\,\text{nm}/$V, cf.\ main text.}
\label{fig:width_voltage}
\end{figure}

\begin{figure}
\includegraphics[width=1.0\columnwidth]{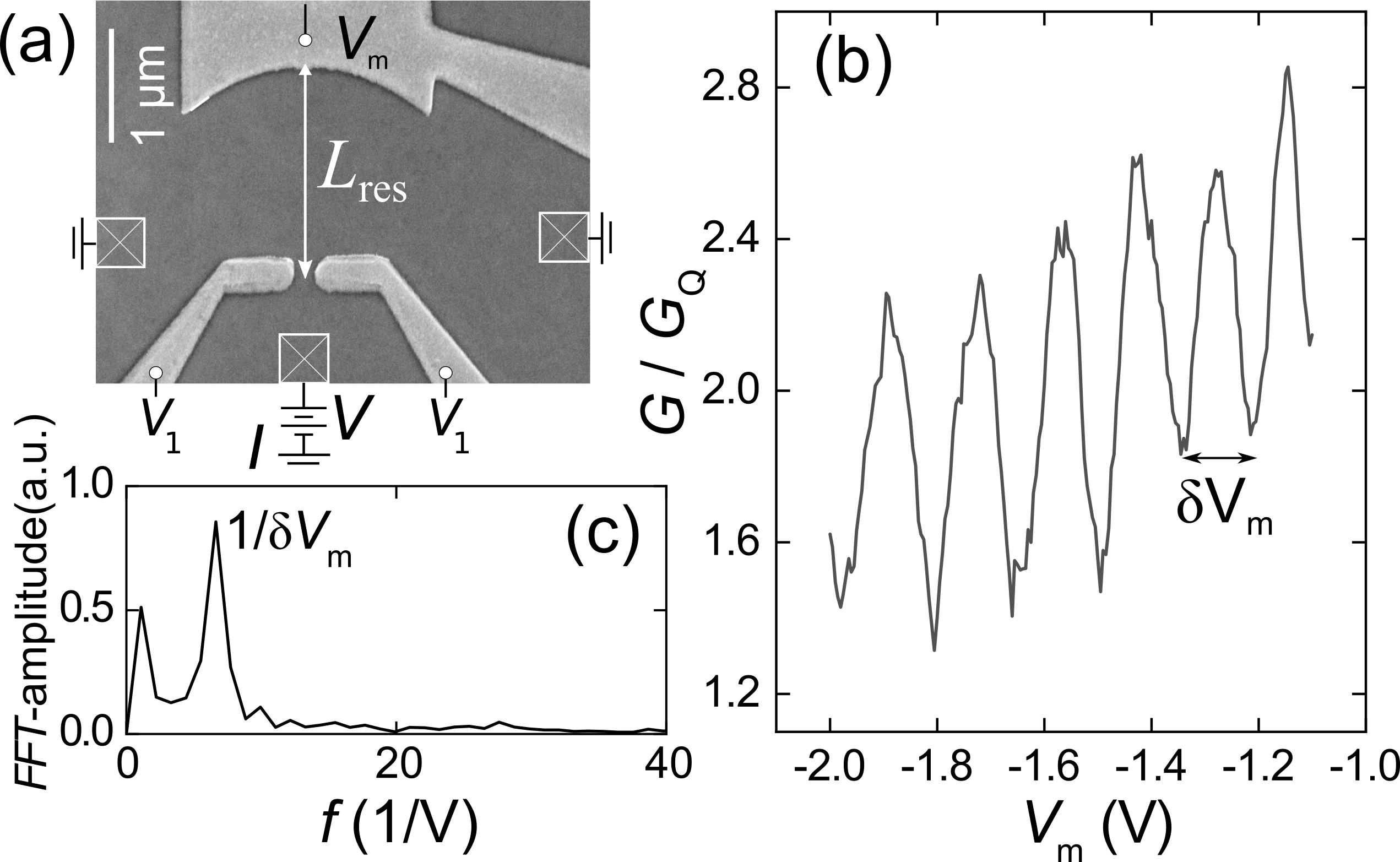}
\caption{(a) QPC nominally identical to QPC$_1$ for $N=4$ coupled to a hemispherical mirror defined by a negative gate voltage $V_\text m$.
(b) Conductance of the QPC as a function of $V_\text m$.
(c) Fourier transform of the conductance. From the peak value, we determine the period $\delta V_\text m\simeq150\,$mV of the oscillations in panel (b).
}
\label{fig:interferometry}
\end{figure}

In \fig{fig:quantumwell-parameter}a we have presented the width of the hard-wall potential $W(N)$. In \fig{fig:width_voltage}{}
we plot $W(\vg)$ for QPC$_1$. Next, we compare this result with the dependence of the depletion region of a gate voltage using a different sample on the same wafer material. The sample shown in \fig{fig:interferometry}a
contains a QPC nominally identical to QPC$_1$ and a hemispherical mirror gate. The two samples have been prepared in parallel and on the same wafer. In \fig{fig:interferometry}b we present the conductance of the QPC as a function of the voltage applied to the mirror gate. The bare conductance (without mirror) is $G=4\gq$. However, with the 2DES below the mirror gate depleted it is reduced roughly by a factor 2, because of enhanced back scattering through the QPC. At the same time $G(V_\text m)$ oscillates with a visibility of 40\,\%. Both, the conductance reduction and oscillation are related to the formation of localized modes inside the hemispherical resonator. The oscillation can be interpreted in analogy to the oscillations of the standing wave in a Fabry-P\'{e}rot resonator, while here, the coherent electrons generate the standing wave. By increasing the gate voltage $V_\text m$ we decrease the area of 2DES depleted next to the mirror gate and thereby increase the length of the resonator (the distance between the QPC and the mirror). Per period of the conductance oscillation the length of the resonator is reduced by half of the Fermi wavelength $\text d L_\text{res}/\text d V_\text m=0.5\lambda_\text F/\delta V_\text m$ with the resonator length $L_\text{res}$. We determine the averaged period from the fast Fourier transform of the oscillation, cf.\ \fig{fig:interferometry}c, and find $\delta V_\text m\simeq150\,$mV. With $\lambda_\text{F}=45\,$nm we finally estimate the rate of the depletion length reduction as  $\text d L_\text{res}/\text d V_\text m=150\,\text{nm}/$V. Changing the voltage applied to the QPC gates instead of the mirror gates results in the same conclusion while in this case the interference pattern appears on top of the QPC pinch-off curve.


To estimate the depletion of the electron system between the QPC gates we have to add up the electric fields of both gates. Based on the fact that the same voltage is applied to both gates and the distance between gates is more than two times larger than the distance between each gate and the 2DES we neglect any influences that a gate has on the depletion caused by the other gate. From the slope of the red line in \fig{fig:width_voltage}{} we find the dependence of the width of the hard-wall potential $W(\vg)$ as a function of gate voltage to be $\text d W/\text d\vg\simeq300\,\text{nm}/$V, twice as large as the effect of a single mirror gate. This finding supports the applicability of the hard-wall model for QPCs with $N\ge4$.

\section{Acknowledgments}
We thank Philipp Altpeter for technical support and are grateful for financial support from the DFG via Grant No. LU 819/11-1. M.G. acknowledges support by project A03 of the CRC-TR 183. A.D.W. acknowledges gratefully support of DFG-TRR160, BMBF - Q.Link.X 16KIS0867 and the DFH/UFA CDFA-05-06.

M. Geier and J. Freudenfeld contributed equally to this work.

\bibliography{literature,../../../zitate/zitate}

\end{document}